\documentclass[12pt]{article}
\usepackage{amsmath}
\usepackage{blindtext} 
\usepackage{subfig}
\usepackage{graphicx}

\usepackage[sc]{mathpazo} % Use the Palatino font
\usepackage[T1]{fontenc} % Use 8-bit encoding that has 256 glyphs
\usepackage{microtype} % Slightly tweak font spacing for aesthetics

\usepackage[english]{babel} % Language hyphenation and typographical rules

\usepackage[hmarginratio=1:1,top=32mm,columnsep=20pt]{geometry} 
\usepackage{booktabs} % Horizontal rules in tables

\usepackage{lettrine} 

\usepackage{enumitem} % Customized lists
\setlist[itemize]{noitemsep} % Make itemize lists more compact

\usepackage{abstract} % Allows abstract customization
\usepackage{titlesec} % Allows customization of titles
\renewcommand\thesection{\Roman{section}} % Roman numerals for the sections
\renewcommand\thesubsection{\roman{subsection}} % roman numerals for subsections
\titleformat{\section}[block]{\large\scshape\centering}{\thesection.}{1em}{} 
\titleformat{\subsection}[block]{\large}{\thesubsection.}{1em}{} 

\usepackage{fancyhdr} % Headers and footers
\usepackage{titling} % Customizing the title section

\usepackage{hyperref} % For hyperlinks in the PDF

\title{Innermost stable circular orbits of a Kerr-like Metric with Quadrupole} 
\author{%
\textsc{Fabi\'an Chaverri-Miranda} \\[1ex] % Your name
\textsc{Francisco Frutos-Alfaro} \\[1ex] % Your name
\textsc{Diego Solano-Alfaro} \\[1ex]
\textsc{Pedro G\'omez-Ovares} \\[1ex] % Your name
\textsc{Andr\'e Oliva} \\[1ex]
\normalsize{School of Physics and 
Space Research Center of the University of Costa Rica}} 
\date{\today} % Leave empty to omit a date
\renewcommand{\maketitlehookd}{%
\begin{abstract}
\noindent 
This paper presents results for the innermost stable circular orbit in a Kerr-like spacetime. The metric employed is an approximation that combines the Kerr metric with the Erez-Rosen metric, expanded in a Taylor series. Consequently, this spacetime incorporates three relativistic multipole moments: mass, spin, and quadrupole moment. Our derivation builds upon the analysis conducted by Chandrasekhar for the Kerr metric. Utilizing the Euler-Lagrange method and Hamiltonian dynamics, we define an effective potential for the radial coordinate. This equation can be used to measure the mass quadrupole through observational methods, as it yields a quadratic polynomial for the quadrupole moment. As anticipated, the limiting cases of this equation correspond to the established cases of Kerr and Schwarzschild spacetimes.
\end{abstract}
}

\begin{document}

% Print the title
\maketitle

%-------------------------------------------------------------------------------
%	ARTICLE CONTENTS
%-------------------------------------------------------------------------------

\section{Introduction}

\noindent 
In classical mechanics, the orbit of a test particle around a massive object can be arbitrary, as the effective potential attains a minimum for any angular momentum value. However, in specific cases, such as the geodesic orbits in a Newtonian gravitational field analogue of Kerr black hole (Euler problem), the system of differential equations is integrable, similar to the dynamics around a Kerr black hole \cite{eleni}. In general relativity, the effective potential in the Schwarzschild metric has two extrema. At the minimum angular momentum, these extrema merge into a single radius: the innermost stable circular orbit (ISCO) \cite{tsupko2016}. The ISCO represents the smallest stable circular orbit for a test particle, often interpreted as the inner boundary of the accretion disk surrounding compact objects.

\noindent
Naturally, the rotation of the central body influences the particle's motion, which is why orbits differ between rotating and non-rotating black holes \cite{jefremov2015}. Another important feature of a compact object is its mass quadrupole moment. It also affects the orbital dynamics of massive and massless particles around a neutron star \cite{of2021}. Yagi and Yunes discover universal relations connecting the moment of inertia, the Love number, and the quadrupole moment, which are unaffected by the internal structure of neutron stars and quark stars. These relations can be employed, for example, to infer the deformation of these compact objects from observations of their moment of inertia \cite{yagiyunes}. The mass quadrupole moment is a quantification of such deformations, which are relevant for rapidly rotating neutron stars \cite{quevedo2016}. Chakrabarti et al., Pappas, Cipolleta et al., and Luk et al. have identified other interesting features for the ISCO of neutron stars, such as a relation that can be used to measure or constrain the radius or to improve the precision of the radius measurement, the constraint on the equation of state (EoS) of matter inside neutron stars. Moreover, as long as there is an ISCO, relations to associate this radius and the orbital frequency with the spin frequency and mass of neutron stars could be found \cite{chakrabarti,pappas2015,cipolletta2017,luk2018}. Furthermore, studying wave emissions and chaotic particle trajectories near compact objects would be particularly insightful for understanding potential deviations from general relativity, as well as the behavior of accretion dynamics and jet formation \cite{DK}. 

\noindent
For many exact solutions of the Einstein field equations, deriving analytical expressions for ISCO radii and frequencies is challenging \cite{berti2004app,pachon2006,sanabria2010}. Even solving the geodesic equations or deducing geodesic properties analytically can be cumbersome. Since it provides information about the spacetime and background geometry near the black hole, understanding the ISCO for black holes is crucial \cite{quevedo2016,pradhan2012}. Additionally, it provides a first approximation to the inner radius of an accretion disk that surrounds the black hole. According to the no-hair theorem, all stationary black hole metrics, such as the Kerr-Newman metric outside the black hole horizon, are fully characterized by three parameters: mass, spin parameter, and electric charge. However, for neutron stars, parameters such as deformation or mass quadrupole and magnetic dipole come into play. For this reason, from the EoS, one obtains the necessary information to fully describe the surrounding spacetime. In this work, we focus on compact objects characterized by 
 mass, spin parameter, and mass quadrupole \cite{shibata1998}.

\noindent
In this paper, we analyze the ISCO for a Kerr-like metric with mass quadrupole (KLMQ). A feature of this metric is that it reduces to Kerr and Hartle-Thorne (HT) spacetimes in specific limits \cite{frutos2013,frutos2014}. By matching the KLMQ with the HT metric, it is possible to find an inner solution. Furthermore, the ISCO equation for the HT metric has been previously derived \cite{berti2004isco,kuantay2015}. Since the KLMQ was derived from the Kerr metric, we expect the ISCO radius, energy, and angular momentum to converge with the known results for Schwarzschild and Kerr black holes \cite{frutos2015,frutos2016}.

\noindent
This paper is organized as follows: Section 2 introduces the KLMQ. Section 3 presents a detailed ISCO calculation using the Euler-Lagrange method, as developed by Chandrasekhar \cite{chandrasekhar1998}. In Section 4, by means of a REDUCE program \cite{REDUCE}, the ISCO equation is compared with known solutions for Kerr and Schwarzschild black holes. Finally, Section 5 summarizes and discusses the results.

%------------------------------------------------

\section{The Kerr-like Metric}

\noindent 
The KLMQ describes the spacetime of a massive, rotating, and deformed object. 
It has three parameters: the mass of the object, $ M $ , the rotation parameter, $ a $ and the quadrupole parameter, $ q $. This metric was generated including the quadrupole moment as a pertubation valid up to the second order in $q$ \cite{frutos2015}. It is an approximate solution of the Einstein field equations and has the following form

\begin{align}
\label{space}
d s^{2} & = g_{tt} d t^{2} + 2 g_{t\phi} d t d \phi + g_{rr} d r^{2} 
+ g_{\theta\theta} d \theta^{2} + g_{\phi\phi} d \phi^{2} ,
\end{align}

\noindent
where the components of the metric are

\begin{align}
\label{met1} 
g_{tt} & = - \dfrac{{\rm e}^{-2\psi}}{\Sigma^{2}}\left[\Delta 
- a^{2} \sin^{2}{\theta} \right] \nonumber \\
g_{t\phi} & = -\dfrac{2Jr}{\Sigma^{2}} \sin^{2}{\theta} \nonumber \\
g_{rr} & = \Sigma^{2} \dfrac{{\rm e}^{2\chi}}{\Delta} \\
g_{\theta\theta} & = \Sigma^{2} {\rm e}^{2\chi} \nonumber \\
g_{\phi\phi} & = \dfrac{{\rm e}^{2\psi}}{\Sigma^{2}}
\left[\left(r^{2} + a^{2}\right)^{2} - a^{2} \Delta \sin^{2}{\theta} \right] 
\sin^{2}{\theta} , \nonumber \\
\end{align}

\noindent
with $ J = M a $,  $ \Sigma^{2} = r^{2} + a^{2}\cos^{2}{\theta} $ and $ \Delta = r^{2} - 2 M r + a^{2} $. The exponents $ \psi $ and $ \chi $ are given by

\begin{align}
\label{exps1}
\psi & = \dfrac{q}{r^{3}}P_{2} + 3 \dfrac{M q}{r^{4}}P_{2} \\
\chi & = \dfrac{q}{r^{3}}P_{2} + \dfrac{1}{3} \dfrac{M q}{r^{4}} 
\left(5 P_{2}^{2} + 5 P_{2} - 1 \right) \nonumber 
+ \dfrac{1}{9} \dfrac{q^{2}}{r^{6}} 
\left(25 P_{2}^{3} - 21 P_{2}^{2} - 6 P_{2} + 2 \right) . \nonumber
\end{align}

\noindent
The function $ P_2 $ is a Legendre polynomial, i. e. $ P_{2} = (3 \cos^{2}{\theta} - 1)/2 $. 

\noindent
As limiting cases this spacetime contains the Kerr metric ($ q = 0 $), the Erez-Rosen metric expanded in Taylor series up to the second order in $ q $ ($ a = 0 $), 
the Lense-Thirring metric (slow rotation, $ a^2 = 0 $), and the Schwarzschild metric ($ q = a = 0 $).

%------------------------------------------------

%\newpage

\section{Deriving the ISCO}

\noindent
The method devised by Chandrasekhar is employed to obtain the ISCO equation \cite{chandrasekhar1998}. The Lagrangian is defined by

\begin{align}
\label{lag1}
L & = \dfrac{\mu}{2} \left(\dfrac{ds}{d\lambda} \right)^2 \nonumber \\
  & = \dfrac{\mu}{2}(g_{tt}\dot{t}^{2} + 2 g_{t\phi} \dot{t} \dot{\phi} 
+ g_{rr}\dot{r}^{2} + g_{\theta\theta}\dot{\theta}^{2} + g_{\phi\phi}\dot{\phi}^{2}) .
\end{align}

\noindent
The dot over the variables $ t, \, r, \, \theta $ and $ \phi $ means derivative with respect to $ \lambda $, the affine parameter. 
To determine the ISCO of a test particle in the plane, one sets $ \dot{\theta} = 0 $ and $ \theta = {\pi}/{2} $. This leaves the Lagrangian as follows

\begin{align}
\label{lag2}
L = \dfrac{\mu}{2}\left(g_{tt} \dot{t}^{2} + 2 g_{t\phi} \dot{t} \dot{\phi} 
+ g_{rr} \dot{r}^{2} + g_{\phi\phi} \dot{\phi}^{2} \right) ,  
\end{align}

\noindent
where the components of the metric becomes

\begin{align}
\label{met2} 
g_{tt} & = \dfrac{e^{- 2 \psi'}}{r^2} \left[2Mr-r^{2} \right] 
\nonumber \\
g_{t\phi} & = - \dfrac{2J}{r} \nonumber \\
g_{rr} & = \dfrac{r^{2} {\rm e}^{2\chi'}}{\Delta}  \\
g_{\theta\theta} & = r^{2} {\rm e}^{2\chi'} \nonumber \\
g_{\phi\phi} &=\dfrac{{\rm e}^{2\psi'}}{r^2}\left[r^{4} + 2M r a^{2} 
+ r^{2} a^{2}\right] , \nonumber 
\end{align}

\noindent
with the exponents $ \psi $ and $ \chi $ of equation (\ref{exps1}) are reduced using Taylor series to

\begin{align}
\label{exps2} 
\psi' & = - \frac{1}{2} \dfrac{q}{r^{3}} - \frac{3}{2} \dfrac{M q}{r^{4}} \\
\chi' & = - \frac{1}{2} \dfrac{q}{r^{3}} - \frac{3}{4} \dfrac{M q}{r^{4}} 
- \frac{3}{8} \dfrac{q^{2}}{r^{6}} . \nonumber
\end{align}

\noindent
The momenta of the three remaining variables are

\begin{align}
\label{moms}
p_{t} & = \mu \left(g_{tt}\dot{t} + g_{t\phi}\dot{\phi}\right) = - E \nonumber \\
p_{r} & = \mu g_{rr}\dot{r} \\
p_{\phi} & = \mu \left(g_{t\phi}\dot{t}+g_{\phi\phi}\dot{\phi}\right) = L_{z} , 
\nonumber
\end{align}

\noindent
where $ \mu = 1 $, $ \rho^{2} = - g_{tt} g_{\phi\phi} + g_{t\phi}^{2} $, $ E $ and $ L_{z} $ are constants of motion that represent the energy and the angular momentum. 
Solving for $ \dot{t} $ and $ \dot{\phi} $ leads to

\begin{align}
\label{vars}
\dot{t} & = \dfrac{1}{\rho^{2}} \left(E g_{\phi\phi} + L_{z} g_{t\phi} \right) \\
\dot{\phi} & = - \dfrac{1}{\rho^{2}} \left(L_{z} g_{tt} + E g_{t\phi} \right) . 
\nonumber
\end{align}

\noindent
The Hamiltonian is defined by

\begin{align}
\label{ham}
H & = \dfrac{1}{2}\left(- E \dot{t} + g_{rr} \dot{r}^{2} 
+ L_{z}\dot{\phi}\right) = \varepsilon \\
& = \dfrac{1}{2} \bigg[ - \dfrac{1}{\rho^{2}}
\left(E^{2} g_{\phi\phi} + 2 E L_{z} g_{t\phi} + L_{z}^{2} g_{tt}\right) 
+ g_{rr} \dot{r}^{2} \bigg] , \nonumber 
\end{align}

\noindent
where $ \varepsilon = -1 $ for time-like geodesics, $ \varepsilon = 0 $ for light-like geodesics and $ \varepsilon = 1 $ for space-like geodesics.

\noindent
We define the effective potential $ V_{\rm eff} $ as

\begin{align}
\label{Vef}
V_{\rm eff} & = - \dfrac{2\varepsilon}{g_{rr}} 
- \dfrac{1}{\rho^{2} g_{rr}}(E^{2} g_{\phi\phi} 
+ 2 E L_{z} g_{t\phi} + L_{z}^{2} g_{tt}) .
\end{align}

\noindent
Using $ u = 1/r $ and the equations (\ref{met2}) and (\ref{exps2}), the effective potential in (\ref{Vef}) can be reduced to

\begin{align}
\label{peff} 
V_{\rm eff} & = - 2\varepsilon \bigg(1 - 2 M u + a^{2} u^{2} + q u^{3} 
- \frac{1}{2}{M q u^{4}} + \frac{5}{4} {q^{2} u^{6}} \bigg) \nonumber \\
& + L_{z}^{2} \left(u^{2} + 2 q u^{5} + \frac{1}{2} {M q u^{6}} 
+\frac{11}{4} {q^{2} u^{8}} \right) - 2 M u^{3}(L_{z} - E a)^{2} \nonumber \\
& - E^{2} \left(1 + a^{2} u^{2} - \frac{3}{4} {M q u^{4}} 
+ \frac{3}{4} {q^{2} u^{6}} \right) 
\end{align}

\noindent
The latter expression has to be differentiated twice to find the values of $ r $ where the orbit is stable:

\begin{align}
\label{dpeff}
\dfrac{dV_{\rm eff}}{du} & = - 2 \varepsilon \bigg(- 2 M + 2 a^{2} u 
+ 3 q u^{2} - 2 M q u^{3} + \frac{15}{2} {q^{2} u^{5}} \bigg) \nonumber \\
& + L_{z}^{2} \left(2 u + 10 q u^{4} + 3 M q u^{5} + 22 q^{2} u^{7} \right) 
- 6 M u^{2} \left(L_{z} - E a \right)^{2} \nonumber \\
& - E^{2} \left(2 a^{2} u - 6 M q u^{3} + \frac{9}{2} {q^{2} u^{5}} 
\right) 
\end{align}

\begin{align} 
\label{d2peff}
\dfrac{d^{2}V_{\rm eff}}{du^{2}} 
& = - 2 \varepsilon \left(2 a^{2} + 6 q u - 6 M q u^{2} 
+ \frac{75}{2} {q^{2} u^{4}} \right) \nonumber \\
& + L_{z}^{2} \left(2 + 40 q u^{3} + 15 M q u^{4} + 154 q^{2} u^{6} \right) 
- 12 M u \left(L_{z} - E a \right)^{2} \nonumber \\
& - E^{2} \left(2 a^{2} - 18 M q u^{2} + \frac{45}{2} {q^{2} u^{4}} \right) 
\end{align}

\noindent
Now, we rewrite these equations using $ x = a E - L_{z} $ as follows

\begin{align}
{\tilde V}_{\rm eff} & = V_{\rm eff} \left[1 - 2 q u^3 + (2 q u^3)^2 \right]  \nonumber \\
& = \varepsilon \bigg(2 - 4 M u + 2 a^2 u^2 - 2 q u^3 
+ 7 M q u^4 + \dfrac{13}{2} q^2 u^6 \bigg) \nonumber \\
& + E^2 \bigg(- 1 + 2 q u^3 + \dfrac{3}{2} M q u^4 
- \dfrac{19}{4} q^2 u^6 \bigg) - 2 E x a u^2 \nonumber \\
& + x^2 u^2 \bigg(1 - 2 M u + \dfrac{9}{2} M q u^4 + \dfrac{11}{4} q^2 u^6 
\bigg) = 0 , \label{pefftecho} \\
{{\tilde V}_{\rm eff}}' & = 
\dfrac{{d} {V}_{\rm eff}}{{d} {u}} \left[1 - 5 q u^3 + (5 q u^3)^2 \right]  
\nonumber \\
& = \varepsilon\big(- 4 M + 4 a^2 u + 6 q u^2 + 16 M q u^3 
- 15 q^2 u^5\big) \nonumber \\
& + E^2 q u^3 \left(6 M - \dfrac{9}{2} q u^2 \right) 
- 4 E x a u \nonumber \\
& + x^2 u \big(2 - 6 M u + 33 M q u^4 + 22 q^2 u^6 \big) = 0 , 
\label{dpefftecho} \\
{{\tilde V}_{\rm eff}}'' & = 
\dfrac{{d}^2 V_{\rm eff}}{{d} {u}^2} \left[1 - 20 q u^3 + (20 q u^3)^2 \right] 
\nonumber \\
& =\varepsilon (4 a^2 + 12 q u - 12 M q u^2 - 165 q^2 u^4) \nonumber \\
& + E^2 q u^2 \left(18 M - \dfrac{45}{2} q u^2 \right) - 4 E x a \nonumber \\
& + x^2 \big(2 - 12 M u + 255 M q u^4 + 154 q^2 u^6\big) = 0 , 
\label{d2pefftecho}
\end{align}

\noindent
where the expressions are set to zero, because we are interested in determining the ISCO equation.

\noindent
From (\ref{pefftecho}) and (\ref{dpefftecho}), $ E^2 $ is found

\begin{align}
\label{E2}
E^2 & = \varepsilon(2 - 2 M u - q u^3 - 8 M q u^4 + 7 q^2 u^6) \nonumber \\
& + x^2 u^3 \left(M - 10 M q u^3 - \frac{33}{4} q^2 u^5 \right) . 
\end{align}

\noindent
A fourth order polynomial for $ x $ is obtained from (\ref{dpefftecho}) and (\ref{E2})

\begin{align} 
{\cal A} x^4 + 2 {\cal B} x^2 + {\cal C} = 0 , 
\end{align} 

\noindent
where

\begin{align} 
{\cal A} & = u^2 (1 - 6 M u + 9 M^2 u^2 - 4 M a^2 u^3 
+ 33 M q u^4 + 22 q^2 u^6) \nonumber \\
{\cal B} & = - 2\varepsilon u \bigg(M + (a^2 - 3 M^2) u 
\nonumber \\
&+ \left(M a^2 u^2 - \frac{3}{2} q \right) u^2 - \frac{5}{2} M q u^3 
+ 6 q^2 u^5\bigg) \nonumber \\
{\cal C} & = \varepsilon^{2} \big(4 M^2 - 8 M a^2 u + 4 a^4 u^2 
- 12 M q u^2 + 9 q^2 u^4 \big).
\end{align}

\noindent
The solution for $ x^2 $ is given by

\begin{align}
\label{x2}
x^2 & =\frac{2\varepsilon}{u Z_{\mp}} 
\bigg[\left(a \sqrt{u} \pm \sqrt{M} \right)^2 
- \frac{3}{2} q u^2 - 7 M q u^3 + 6 q^2 u^5 \bigg] ,
\end{align}

\noindent
with

\begin{align}
Z_{\pm} & = \left(1 - 3 M u + \frac{33}{2} M q u^4 + 11 q^2 u^6 \right) 
\pm 2 a u \sqrt{M u} .
\end{align}

\noindent
Inserting (\ref{x2}) in (\ref{E2}) one finds $ E^2 $  

\begin{align}
\label{E2R} 
E^2 & = \frac{2\varepsilon}{Z_{\mp}}
\bigg[(1 - 2 M u) \left(1 - 2 M u \pm 2 a u \sqrt{M u} \right) \\
& + \left(a^2 M - \frac{1}{2} q \right) u^3 + \frac{25}{2} M q u^4 
+ \frac{29}{2} q^2 u^6 \bigg] \nonumber
\end{align}

\noindent
Substituting (\ref{E2R}) and (\ref{x2}) in (\ref{dpeff}), $ L_{z}^{2} $ is determined

\begin{align}
\label{Lz2}
L_z^2 & =\frac{2\varepsilon u}{{\cal A}} \bigg[M - 3 M^2 u 
+ \left(2 M a^2 - \frac{3}{2} q \right) u^2 
+ \left(6 M^2 a^2 - \frac{5}{2} M q \right) u^3 \nonumber \\
& + M \left(a^4 - 12 M^2 a^2\right) u^4 + \left(5 M^2 a^4 + 6 q^2\right) u^5 
\nonumber \\
& \pm 2 M a u \sqrt{M u} \big(a^4 u^4 - 2 M a^2 u^3 
+ 4 a^2 u^2 - 6 M u + 3\big) \bigg] , 
\end{align}

\noindent
where $ {\cal A} = u^2 Z_{+} Z_{-} $.

\noindent
Finally, substituting (\ref{E2R}), (\ref{Lz2}) and (\ref{x2}) in (\ref{d2peff}) 
and changing $ u = 1/r $, the ISCO equation is found

\begin{align}
\label{pisco}
{\cal P} & = M r^5 - 9 M^2 r^4 
+ 3 \left(6 M^3 - M a^2 + \frac{1}{2} q \right) r^3 \nonumber \\
& - \left(7 M^2 a^2 - \frac{29}{2} M q \right) r^2 
- \frac{33}{2} q^2 \pm 6 M a r \sqrt{M r} \Delta = 0 . 
\end{align}

\noindent
To test equation (\ref{pisco}), we introduced values of each parameter that model several neutron star scenarios (see Table 1). Each model was produced with the Rapidly Rotating Neutron Star code ({\text{https://github.com/cgca/rns}}) \cite{RNS}. The first three configurations were produced with the EoS FPS, whereas the latter was produced with the rigid EoS L. BWFX corresponds to the Black Widow pulsar (PSR B1957+20). SHFT represents PSR J1748-2446ad, the fastest known spinning pulsar with a rotation frequency of 716 Hz. Additionally, KAFT and KALN are hypothetical pulsars with rotation frequencies of $1000$ Hz. The results of these tests are presented in Table \ref{data}. For more details on these neutron star models, see \cite{of2021}.

It is important to note that in the BWFX case, due to the two possible sign choices for the $\cal{P}$ function, multiple solutions for the ISCO may exist. In order to keep only the ISCO radii that could be observed in principle, we discard any root smaller than the corresponding radius-to-mass ratio.

\begin{table}[h!]
\centering
\begin{tabular}{c c c c c c}
\hline 
Conf. & $ R (km) $ & $ R/M $ &  $ a/M $ & $ q/M^3 $ & $ R_{ISCO} / M $\\
\hline 
BWFX    & 9.487 & 3.52609 & 0.1913  & 0.06357 & 5.21876, 6.55096 \\
SHFT    & 11.43 & 5.39374 & 0.3306  & 0.44411 & 6.66006 \\
KAFT    & 12.65 & 6.06796 & 0.5257  & 1.17790 & 6.71845 \\
KALN    & 20.05 & 4.99251 & 0.7134  & 1.24577 & 7.44424 \\
\hline
\end{tabular}
\caption{Radius, mass, parameters and ISCO for different configurations.
The first configuration has two possible ISCO radii for corotating and counter-rotating orbits.}
\label{data}
\end{table}

\begin{figure*}[h!]
\centering
\subfloat(a){\includegraphics[width=7cm]{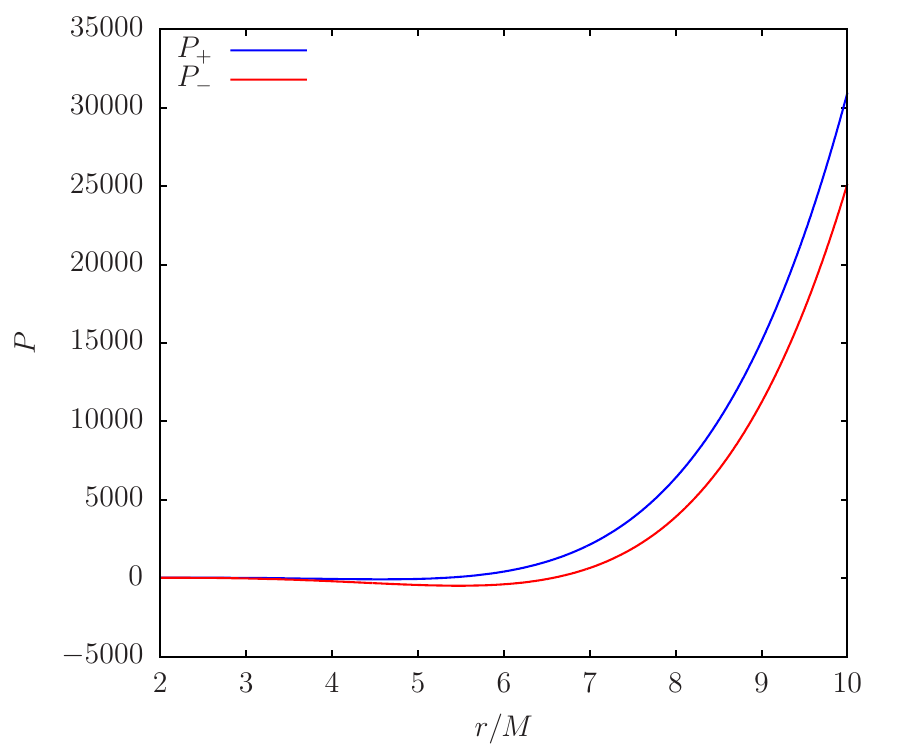}} 
\subfloat(b){\includegraphics[width=7cm]{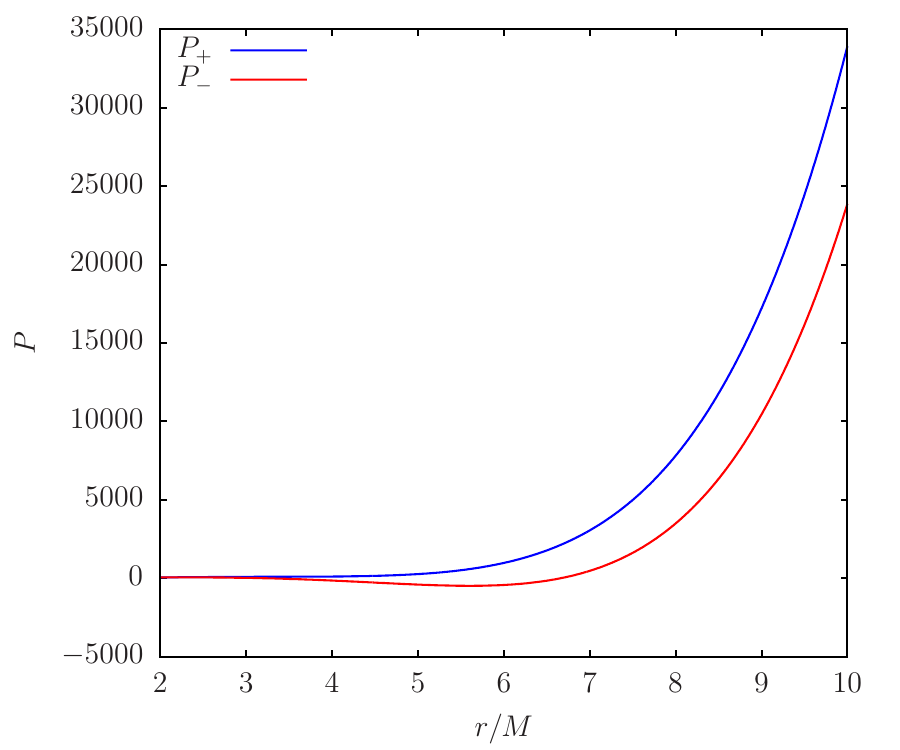}} \\
\subfloat(c){\includegraphics[width=7cm]{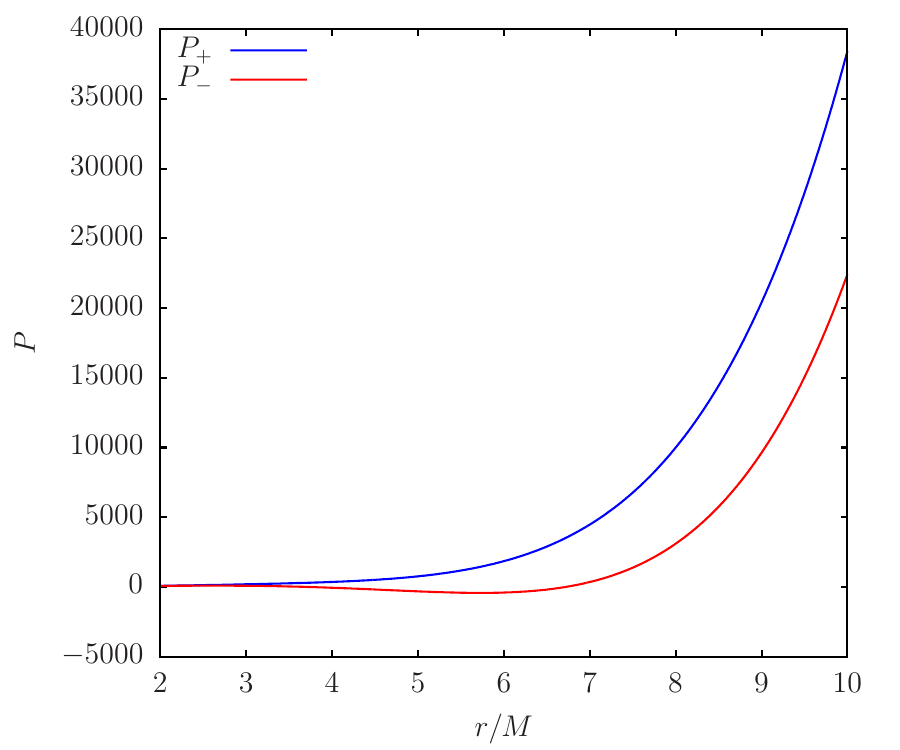}} 
\subfloat(d){\includegraphics[width=7cm]{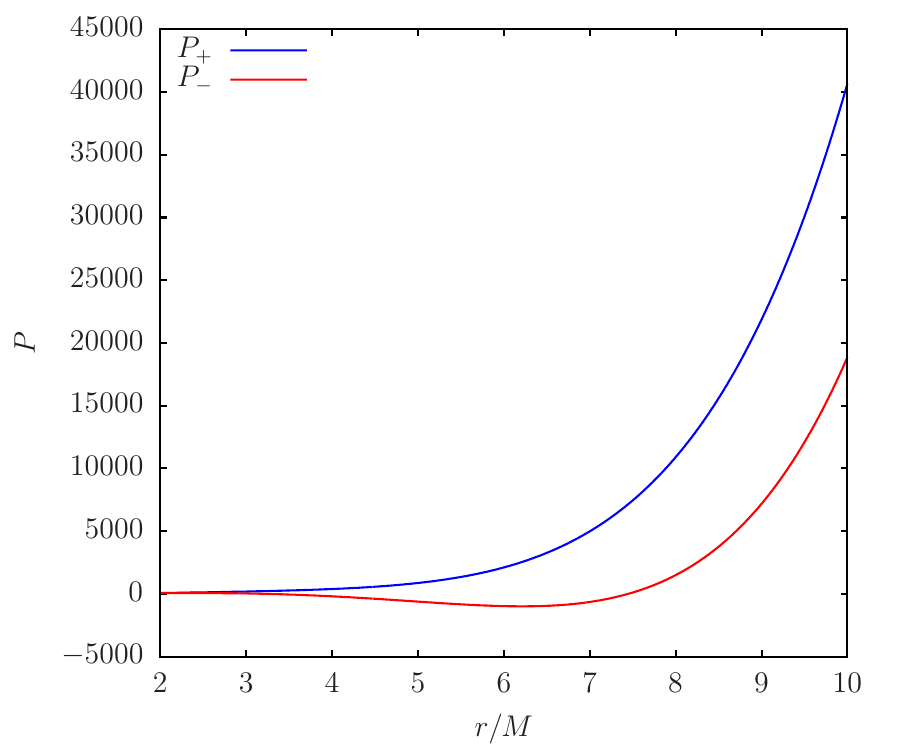}}
\caption{$\cal{P}$ vs. radius for configurations BFXT (a), SHFT (b), KAFT (C), and KALN (d).}
\label{Pplot}
\end{figure*}

%------------------------------------------------

\section{ISCO mass quadrupole moment}

\begin{figure*}[h!]
\centering
\subfloat(a){\includegraphics[width=7cm]{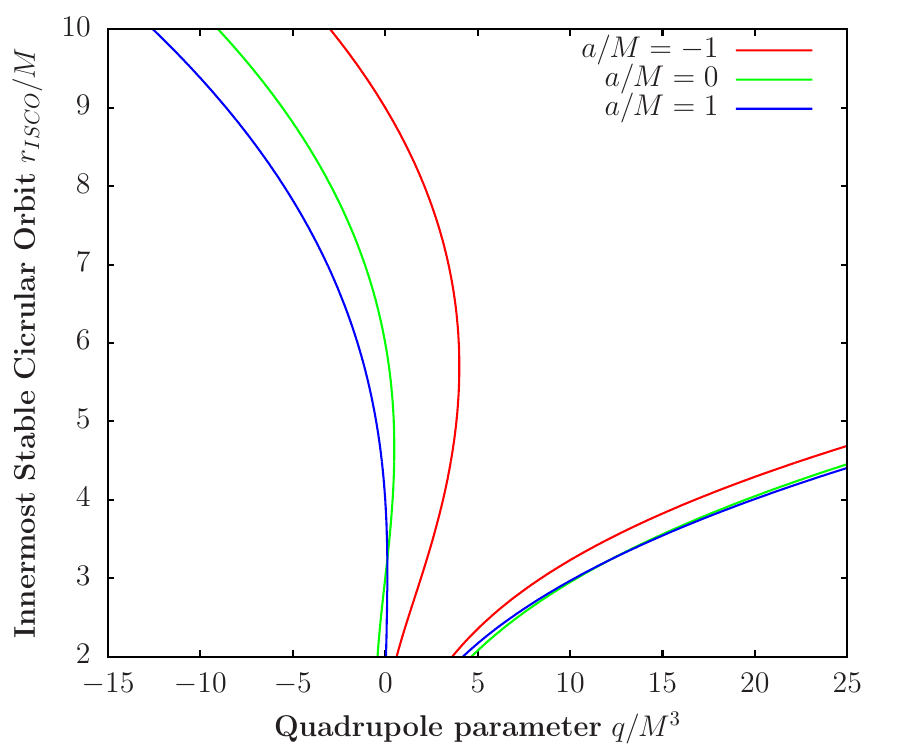}} 
\subfloat(b){\includegraphics[width=7cm]{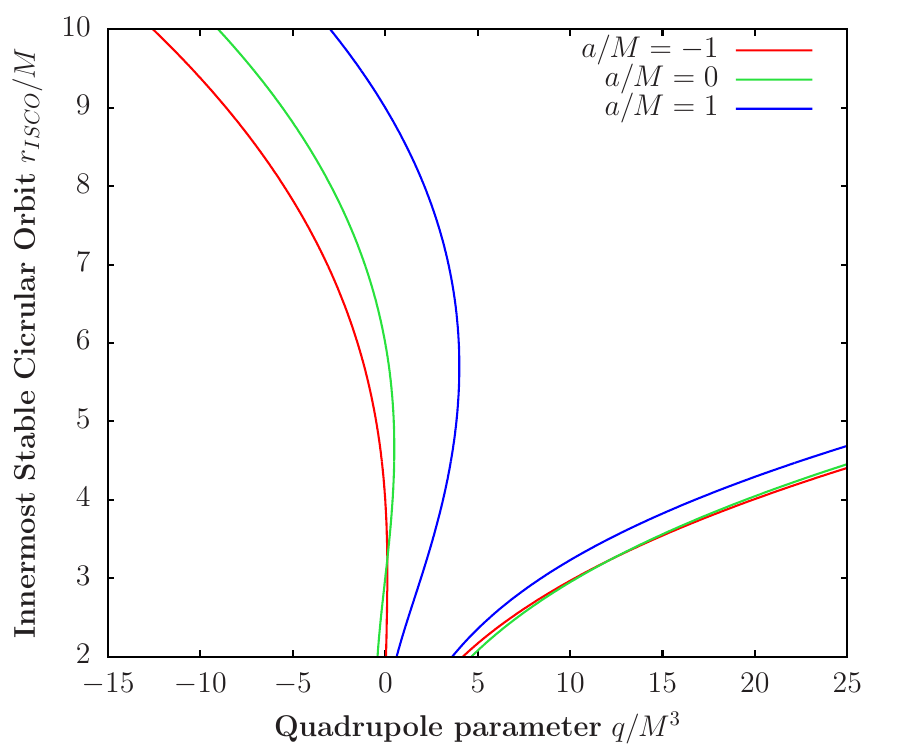}}
\caption{(a) Mass quadrupole solutions for $\cal P_+$ (a) and $\cal P_-$ (b).}
\label{Qplot}
\end{figure*}

Equation (\ref{pisco}) suggests the existence of several values for the ISCO, defined by $\cal P$. Therefore, obtaining the ISCO associated with each quadrupole would be complicated to represent in a simple plot, as the ones shown in Figure \ref{Qplot} for three values of $a$. 
\noindent
For simplicity, we fixed $r_{ISCO}$ and then obtained the possible values for $q$. This resulted in four distinct values for the mass quadrupole corresponding to each ISCO defined by $\cal P$: two values for each sign possibility, $\cal P_+$ and $\cal P_-$, since $q$ depends quadratically on $\cal P$. Lastly, we inverted the axis. 
\noindent
Bare in mind also that there might be a few additional values for the ISCO for each $q$ in the horizontal axis. Plus, we established the correspondent Kerr event horizon as a type of filter for radii that might be too close to the compact object. The total number of roots obtained by solving the $\cal{P}$ function for $r_{ISCO}$ is further analyzed in \cite{effects}.

\section{Limiting Cases}

\noindent
We examine our results for the limit cases contained in the KLMQ metric described in Section III for the possible $ r $ values. For this analysis, we also used a REDUCE program which finds the solutions for equations such as the one obtained in (\ref{pisco}).

\noindent
The first limiting case is when $ q = a = 0 $, which reduces to the known Schwarz\-schild metric case ($ r = 6 M $). For Schwarzschild, the relation found is the following

\begin{align}
\label{piscosch}
{\cal{P}}_{Sch} & = M r^{5}- 9 M^{2} r^{4} + 18 M^{3} r^{3} 
= M r^{3} \left(r - 6 M \right) \left(r - 3 M \right) = 0 .
\end{align}

\noindent
The Schwarzschild case is contained in (\ref{piscosch}), and the values of the energy (\ref{E2R}) and angular momentum (\ref{Lz2}) for this case are also reduced to the literature values \cite{chandrasekhar1998}.

\noindent
The other important case is the Kerr one, for which $ q = 0 $ is set in (\ref{pisco}). 
The ISCO equation for the Kerr metric found by Chandrasekhar and Pradhan \cite{pradhan2012, chandrasekhar1998} is

\begin{align}
\label{kerrchand} 
r^{2} - 6 M r \mp 8 a \sqrt{M r} - 3 a^{2} & = 0 .
\end{align}

\noindent
Squaring the last expression, one gets

\begin{align}
\label{kerrlisco} 
r^{4} - 12 M r^{3} + 6 (6 M^{2} - a^{2}) r^{2} - 28 M a^{2} r + 9 a^{4} & = 0 .
\end{align}

\noindent
The simplification of (\ref{pisco}) is

\begin{align}
\label{piscokerr}
{\cal{P}}_{Kerr} & = M r^{5} - 9 M^{2} r^{4} 
+ 3 \left(6 M^{3} -M a^{2} \right) r^{3} 
- 7 M^{2} a^{2} r^{2} \pm 6 M a r \sqrt{M r}\Delta = 0 . 
\end{align}

\noindent
From last expression, after squaring, we get

\begin{align}
\label{piscokerr2}
(r^{4} - 12 M r^{3} + 6 (6 M^{2} - a^{2}) r^{2} - 28 M a^{2} r + 9 a^{4}) 
& \times & \nonumber \\
(r^{3} - 6 M r^{2} + 9 M^{2} r - 4 M a^{2}) & = 0 .
\end{align}

\noindent
From this, we see that the equation (\ref{kerrchand}) is contained in (\ref{piscokerr}), giving us the solutions for the Kerr case. The energy and angular momentum are

\begin{align}
E & = \sqrt{\frac{1}{Z_{\mp}}} \bigg(1 - 2 M u \mp a u\sqrt{M u} \bigg), \\
x & = - \dfrac{a \sqrt{u} \pm \sqrt{M}}{\sqrt{u Z_{\mp}}}, \nonumber \\
L_{z} &= \mp \sqrt{\dfrac{M}{u Z_{\mp}}}\left(a^{2} u^{2} + 1 
\pm 2 a u \sqrt{M u}\right) .
\end{align}

\noindent
These values for $ E $ and $ L_z $ are exactly the same as the ones determined by Chandrasekhar \cite{chandrasekhar1998} which validates the original equations (\ref{x2}), (\ref{E2}) and (\ref{Lz2}).

\noindent
For $ a = 0 $, an approximation to the ISCO of the Erez-Rosen metric is found. The ISCO formula takes the following form

\begin{align}
\label{piscoer}
{\cal P} & = M r^5 - 9 M^2 r^4 + 3 \left(6 M^3 + \frac{1}{2} q \right) r^3 
+ \frac{29}{2} M q r^2 - \frac{33}{2} q^2 = 0 . 
\end{align}

\noindent
From equations (\ref{pisco}) and (\ref{piscoer}), it is obvious that if the ISCO radius, the mass and rotation parameter are known, 
then a second order polynomial in $q$ can be solved to get an approximate value of $q$ for the compact object.  

%------------------------------------------------

\section{Summary and Conclusions}

\noindent
In this study, we derived the ISCO equation for a Kerr-like metric incorporating a mass quadrupole moment. The derived equation seamlessly reduces to the well-known results for both the Kerr metric ($ q = 0 $) and the Schwarzschild metric ($ q = a = 0 $). Additionally, we were able to obtain an approximation for the ISCO equation in the limit of the Erez-Rosen metric (where $ a = 0 $). Alongside the ISCO equation, we also derived analytical expressions for the energy and angular momentum of the orbiting particle; the latter are expressed as functions of the mass, rotation parameter, and quadrupole moment.

\noindent
In a Schwarzschild black hole, the ISCO radius is $ r_{ISCO} = 6 M $. In the Kerr metric, the ISCO radius depends on the direction of the particle's motion relative to the black hole's rotation. A co-rotating particle has a smaller ISCO ($ r_{ISCO} = M $), while a counter-rotating particle has a larger one ($ r_{ISCO} = 9 M $) \cite{tsupko2016}. By comparison, for the Black Widow pulsar, $ R = 3.53 M $, indicating that some neutron stars are smaller than their ISCO, making the ISCO observable and measurable in orbits of massive particles. Furthermore, this allows for the measurement of parameters like $ q $.

\noindent
A key feature of the ISCO equation is its quadratic dependence on the quadrupole moment $ q $, which greatly simplifies its solution. Given known values of the spin parameter $ a $, the mass $ M $, and the ISCO radius $ r_{ISCO} $, one can, in principle, solve for $ q $. This implies that, the quadrupole moment of a compact object, such as a rotating neutron star or a deformed black hole, can be indirectly measured. In the case of neutron stars where deformation due to rotation plays a significant role, such measurements would provide crucial information about the internal structure and shape of compact objects.

\noindent
Furthermore, Figure \ref{Qplot} provides insight into the relation between the parameters of the KLMQ metric and the ISCO properties. A more detailed analysis of this relationship is provided in \cite{effects}.

\noindent
For future works, we would like to create an outline to measure the quadrupole moment observationally from the ISCO. One of the possibilities is that the transport of material from the disk to the neutron star takes place through the magnetic field lines. Consequently, radius estimations of the magnetospheric accretion are relevant \cite{miller}.   

%-------------------------------------------------------------------------------
%	Bibliography
%-------------------------------------------------------------------------------

%\bibliographystyle{plain}
%\bibliography{References}

%-------------------------------------------------------------------------------

\end{document}